\documentclass[%
 reprint,
 amsmath,amssymb,
 aps,
]{revtex4-2}

\usepackage{graphicx}
\usepackage{dcolumn}
\usepackage{bm}

\begin{document}

\preprint{APS/123-QED}

\title{Specular reflection of polar molecules from a
simple
multi-cylinder
electrostatic mirror:
a method for separating 
BaF
molecules produced in 
a 
buffer-gas-cooled 
laser-ablation 
source 
from other 
ablation products}

\author{H.-M. Yau}
\author{Z. Corriveau}
\author{N. T. McCall}
\author{J. Perez Garcia}
\author{D. Heinrich}
\author{R. L. Lambo}
\author{G. K. Koyanagi}
\author{M. C. George}
\author{C. H. Storry}
\author{M. Horbatsch}
\author{E. A. Hessels}
  \email{hessels@yorku.ca}

\affiliation{%
 Department of Physics and Astronomy,
 York University
}%

\collaboration{EDM$^3$
Collaboration}

\date{\today}

\begin{abstract}
A method for  
specular 
reflection of
polar molecules is proposed. 
Electrostatatic 
potentials and forces
are calculated for a 
low-field-seeking molecule
near 
a series of long cylindrical
electrodes of radius
$r$
with  
dc 
potentials of 
$+V$ 
and
$-V$
applied to alternate electrodes.
A 
center-to-center
separation of 
$2.9\,r$
leads to remarkably flat 
equipotential
surfaces and 
thus to 
a nearly planar mirror for
specular 
reflection of the polar molecules,
with the angle of reflection
equalling the angle of incidence
to an accuracy approaching
a 
microradian.
This 
mirror
can be used to redirect 
cryogenic molecular beams.
Separating  
barium monofluoride 
(BaF) 
molecules 
created in a 
helium-buffer-gas
laser-ablation
source
from other ablation products
is a necessary step to producing
a pure sample of 
matrix-isolated 
BaF,
as is required by the 
EDM$^3$
collaboration for implementing 
a precise measurement of the 
electron electric dipole moment.
The design and modelling for 
the 
BaF 
deflector 
based on this electrode geometry
is presented.
\end{abstract}

\maketitle


\section{\label{sec:intro}Introduction}

The 
Stark
shift of a neutral polar molecule
leads to a potential energy
and force for 
a molecule in an electrostatic field.
These potentials
depend on the 
magnitude of electric 
field, 
and can be 
low- 
or 
high-field 
seeking, 
depending on the 
molecule's
rotational quantum number 
(and its projection) 
and on the magnitude
of the 
electric field.
Deflection of neutral polar
molecules by 
electrostatic 
fields was first 
demonstrated
\cite{wrede1927ablenkung} 
almost a century
ago. 
For 
low-field-seeking 
states,
a region of 
large electric field 
can be used to reflect 
polar molecules.
\cite{wark1992electrostatic,schulz2004microstructured,florez2011electrostatic,Li2020Cold}

In this work,
we present 
a simple geometry that allows
for an 
electrostatic 
polar-molecule
mirror with 
specular
reflections.
Our geometry 
(see 
Fig.~\ref{fig:geometry}a)
consists of 
a series of 
long cylindrical electrodes,
rather than electrodes 
deposited on an insulating substrate
used
\cite{wark1992electrostatic,schulz2004microstructured,florez2011electrostatic}
in previous electrostatic 
mirrors.
The cylindrical  
electrodes allow for
reflections that are nearly 
specular.
The lack of an insulating substrate
avoids possible arcing between 
electrodes, allowing for larger 
voltages to be used. 
This arcing can be of particular
concern for 
cryogenic molecules
created in a 
laser-ablation buffer-gas
source, 
given that
over time
laser-ablation 
products could 
coat the insulating substrate
and create surface pathways for 
arcing.

With the cylinders having 
applied voltages of 
$+V$
and
$-V$
(alternately,
as in 
Fig.~\ref{fig:geometry}a)
and,
with a carefully chosen
separation between the cylinders,
the magnitude of the 
electric field decreases 
exponentially with distance from the 
mirror, 
but shows a remarkably small 
variation within the planes 
parallel to the mirror surface
(Section~\ref{sec:fields}).
This small variation is in contrast
to previous electrostatic mirrors
\cite{wark1992electrostatic,schulz2004microstructured,florez2011electrostatic,Li2020Cold},
and is key to allowing for 
specular 
reflections.

Although the results of this 
work are generally applicable
to any polar molecule,
in 
Section~\ref{sec:shifts},
we specialize to the case
of 
barium monofluoride
(BaF),
since this molecule is 
being employed by the 
EDM$^3$
collaboration 
in an attempt
to 
make an 
ultraprecise 
measurement
\cite{vutha2018orientation}
of the electric
dipole moment of the electron.
For the measurement,
BaF 
molecules 
produced by 
a 
buffer-gas-cooled
laser-ablation 
source
are embedded
in
solid 
argon
\cite{lambo2023calculationAr}
or 
neon
\cite{li2023optical,
lambo2023calculationNe,
corriveau2024matrix}. 
For the precision measurement
it is necessary to have
an uncontaminated solid,
and therefore the 
BaF 
molecules must be 
separated from all of the
other ablation products 
via a deflection.

By careful choice of 
$V$,
one can further improve the
quality of the 
specular 
reflections.
In the 
section~\ref{sec:deflector}, 
we describe how the electrostatic
mirror will be used 
to deflect the 
BaF
molecules
away from the 
other 
ablation products.

\section{\label{sec:fields}Electric Field Profile}

The 
Stark
shift 
experienced by 
a 
slowly-moving 
polar molecule 
depends only on the magnitude of 
the electric field,
since,
even if the 
direction of the
field is varying 
spatially,
the orientation
of the moving molecule follows the 
field direction 
adiabatically
as it moves through the field.
Therefore, 
an ideal electrostatic planar mirror
would have constant 
field magnitudes on planes that 
are parallel to the mirror surface.

\begin{figure}
\centering
\includegraphics[width=3.5in]{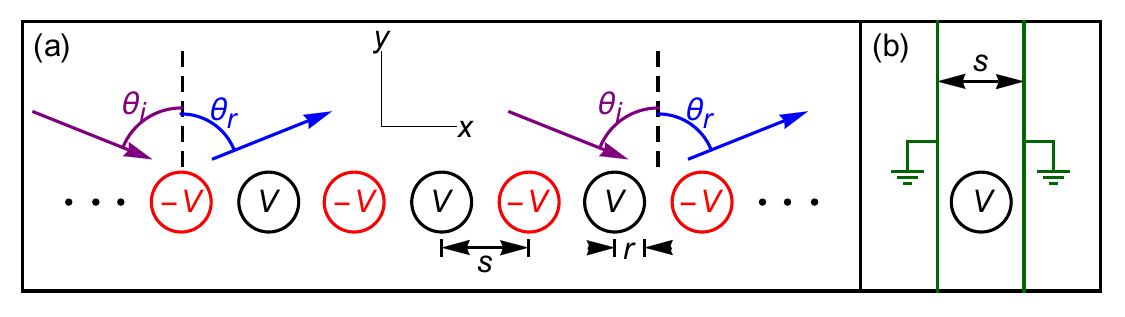}
\caption{
\label{fig:geometry} 
(color online)
The geometry being considered
in this work uses a set of 
long cylindrical 
electrodes of radius
$r$ 
and
center-to-center
separation 
$s$
that are
alternately set 
to 
dc
voltages of 
$+V$
and
$-V$, 
as shown in cross section
in 
panel~(a).
For the separation 
$s$ 
pictured, 
reflections are
almost perfectly 
specular
(with angle of 
reflection, 
$\theta_r$,
equal to the 
angle of incidence,
$\theta_i$,
independent of whether
the incident molecule is 
aimed at the axis of 
one of the cylinders,
(as pictured at the left 
in (a)),
at a point halfway between
two cylinders
(as pictured at the right 
in (a)),
or at any other point,
$x$.
By symmetry, 
the fields can be calculated
from that of a single
cylinder at potential 
$+V$ 
(or
$-V$)
centered between 
two grounded planes
separated by 
$s$,
as shown in 
panel~(b).
} 
\end{figure} 

We consider here the 
simple geometry shown in 
Fig.~\ref{fig:geometry}(a),
in which 
equally spaced long cylinders
of radius 
$r$
have a 
center-to-center
separation 
$s$, 
and have potentials
$+V$
and
$-V$
applied to 
alternate cylinders.
To calculate the 
electrostatic potential
for this problem, 
one can use symmetry to
reduce the problem to 
that of a single cylinder 
at potential 
$V$
centered 
between two 
grounded planes,
as shown in
Fig.~\ref{fig:geometry}(b).
This simplified geometry 
uses the fact that the 
planes bisecting two 
adjacent cylinders have 
an electrostatic potential
of exactly zero. 

The 
electrostatic potential for 
the 
Fig.~\ref{fig:geometry}(b)
geometry can be calculated
in two ways. 
First, 
one can numerically solve 
the 
two-dimensional 
Laplace
equation with a boundary 
condition
of 
$V$
on the circle of radius 
$r$,
centred in a grounded 
rectangle of width
$s$
and height
$h \gg s$.
With an extrapolation of 
$h$ to infinity,
this method allows
for accurate calculations
of the electrostatic
potential.

An analytic solution is also 
possible, 
using line charges and 
the method of images.
An initial line charge
$+\lambda$
at 
$x=y=0$
gives a uniform potential 
on the circle of radius r.
Two image charges
$-\lambda$
are 
needed 
(at 
$x=\pm s$,
$y=0$)
to establish the 
grounded potentials
at 
$x=\pm s/2$
(the green lines in 
Fig.~\ref{fig:geometry}(b)).
With these two image charges,
the 
potential on the
central circle of radius
$r$ is no longer constant.
However, 
this can be 
rectified with two additional
line charges 
$+\lambda$
at 
$x=\pm\, r^2/s$,
$y=0$.
These in turn need image 
charges to reinstate the 
grounded-plane 
boundary conditions.
The result is an infinite 
number of image line charges
(all in the 
$y=0$
plane)
whose locations can be expressed 
as continuing fractions:
\begin{equation}
x_{n_1 n_2 n_3 ... n_N}
=
n_1 s
+
\cfrac{r^2}
{n_2 s
+
\cfrac{r^2}
{n_3 s 
+
\cfrac{r^2}
{
\begin{array}
{@{}c@{}c@{}c@{}}
        n_4 s + {}\\ &\ddots\\ &&{}+ \cfrac{r^2}{n_N s}
      \end{array}
    }}}.
\end{equation}
Here, 
$N$ is an 
integer
$\ge 1$,
and 
the indices
$n_1$
through
$n_N$
can take on all possible 
integer values
except that 
$n_2$
through
$n_N$
must be non-zero.
The line charges all have 
charge densities of 
$\pm \lambda$,
with 
\begin{equation}
\lambda_{n_1 n_2 n_3...n_N}
=
-\lambda
(-1)^
{N+
\sum_{i=1}^{N} n_i}.    
\end{equation}
The electrostatic potential for the 
geometry in 
Fig.~\ref{fig:geometry}
is
then given by
\begin{eqnarray}
\Phi(x,y)
=
\frac
{-\lambda}
{2 \pi \epsilon_0}
\sum_{N=1}^\infty
\,\,
\sum_{n_1,n_2,...,n_N
}
(-1)^
{N+
\sum_{i=1}^{N} n_i}
\nonumber
\\
\times \ln({\sqrt{(x-x_{n_1 n_2  ... n_N})^2+y^2}}),
\end{eqnarray}
where the value of 
$\lambda$
is obtained from
$\Phi(r,0)=V$.
The electric field magnitude is 
\begin{eqnarray}
&&|\vec{E}(x,y)|
=\sqrt{
\Big(
\frac
{\partial \Phi}
{\partial x}
\Big)^2
+
\Big(
\frac
{\partial \Phi}
{\partial y}
\Big)^2
}
\nonumber
\\
&&=
\dfrac
{V}
{\sum\limits_{N=1}^\infty
\,\,
\sum\limits_{n_1,n_2,...,n_N
}
(-1)^
{N+
\sum\limits_{i=1}^{N} n_i}
\ln({\sqrt{x_{n_1 n_2  ... n_N}^2+r^2}})}
\nonumber
\\
&&\times
\Big[
\Big(
\sum_{N=1}^\infty
\,\,
\sum_{n_1,n_2,...,n_N
}
\!\!\!\!\!\!
\frac
{(-1)^
{N+\sum_{i=1}^{N} n_i}
(x-x_{n_1 n_2  ... n_N})}
{\sqrt{(x-x_{n_1 n_2  ... n_N})^2+y^2}}
\Big)^2
\nonumber
\\
&&+
\Big(
\sum_{N=1}^\infty
\,\,
\sum_{n_1,n_2,...,n_N
}
\!
\frac
{(-1)^
{N+\sum_{i=1}^{N} n_i}
y}
{\sqrt{(x-x_{n_1 n_2  ... n_N})^2+y^2}}
\Big)^2
\Big]^{\frac{1}{2}}.
\end{eqnarray}

An approximation to this 
infinite sum can be had by 
a truncation 
that has equal number
of positive 
and
negative 
image line charges 
added inside the
cylinder centred at 
$(x,y)$=$(0,0)$.

\begin{figure}
\centering
\includegraphics[width=3.5in]{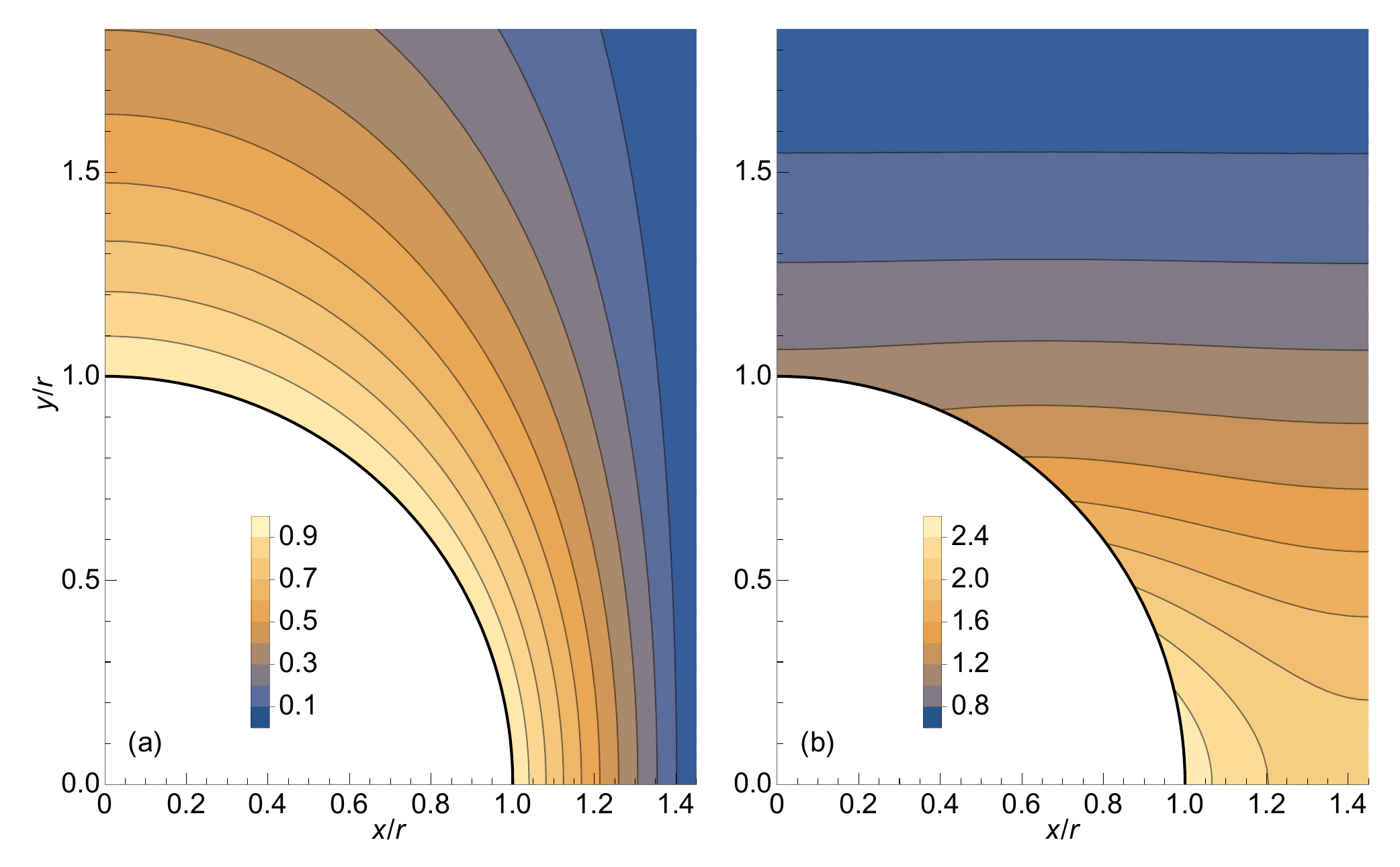}
\caption{
\label{fig:VE} 
(color online)
The electrostatic potential is shown in
(a)
in units of 
$V$, 
and the 
electric field magnitude
is shown in 
(b)
in units of 
$V/r$.
Both plots are for 
$s=2.9 r$,
which leads to nearly 
horizontal contours of 
electric field magnitude
for 
$y>r$,
as seen in 
(b).
} 
\end{figure} 

The electrostatic potential and 
electric field magnitude for the 
case 
$s=2.9 r$
is shown in 
Fig.~\ref{fig:VE}.
For this choice of 
separation,
the electric field magnitude 
contours 
(Fig.~\ref{fig:VE}(b)
for 
$y/r\gtrsim 1.4$)
are nearly flat surfaces.
For  
$s<2.9 r$,
the field magnitude is 
larger at 
$x=s/2$ 
than it is at
$s=0$,
and for 
$s>2.9 r$,
the opposite is true.

\section{\label{sec:shifts}Stark shift and molecular potential energy}

To obtain the potential energy
experienced by a polar molecule
in the electric field of 
Section~\ref{sec:fields},
we need to use the 
Stark
shift
of the molecule.
Any polar molecule
could be used, 
however, 
we focus this work on 
the 
BaF
molecule,
since 
the
BaF
molecule 
is being used 
\cite{vutha2018orientation,
corriveau2024matrix}
by 
the 
EDM$^3$
collaboration.

\begin{figure}
\centering
\includegraphics[width=3.0in]{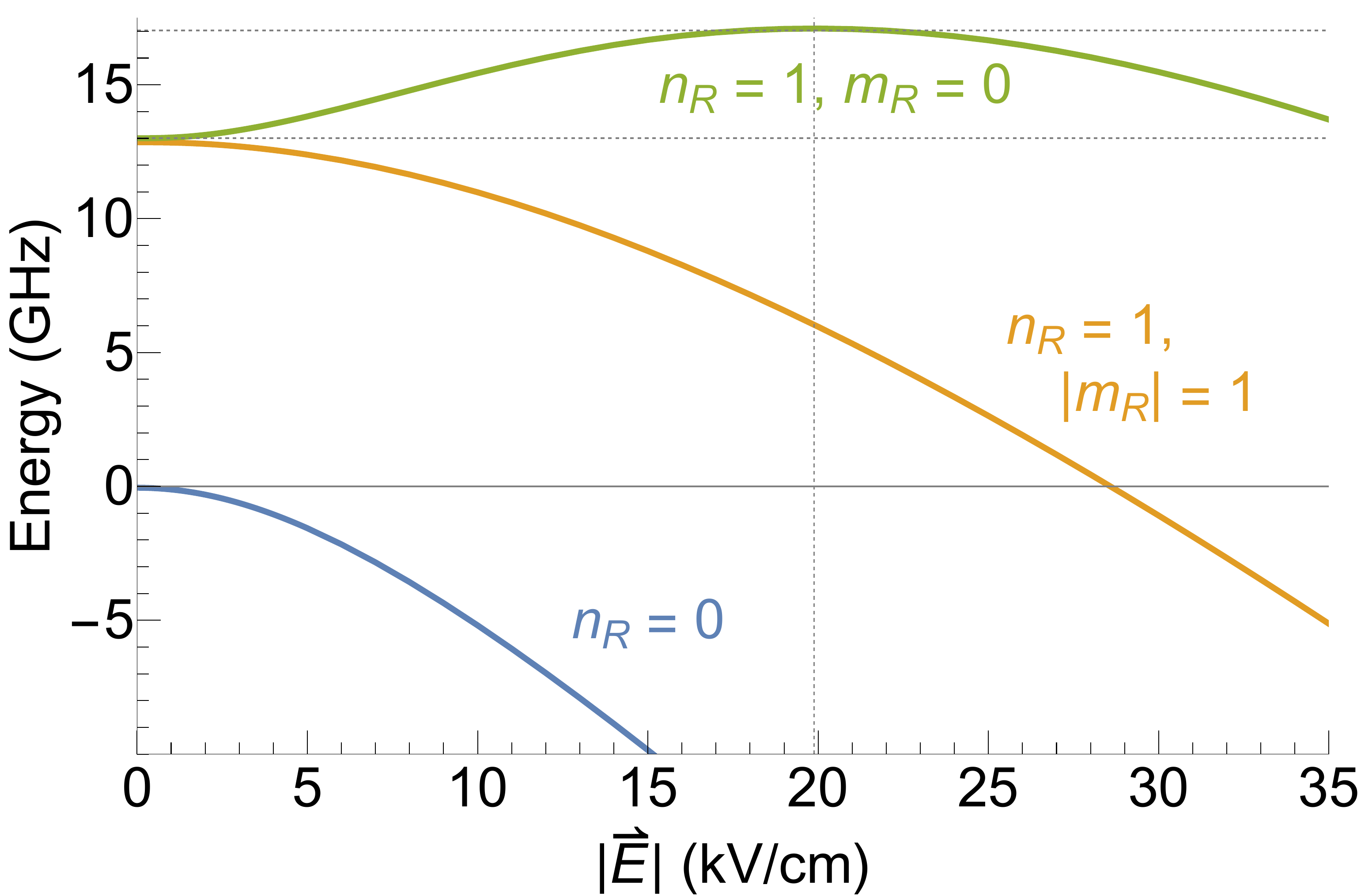}
\caption{
\label{fig:StarkShift} 
(color online)
The 
Stark 
shift of the 
lowest rotational levels
of 
the 
ground electronic
and vibrational state
of 
the
BaF 
molecule.
Molecules in the
$n_{\rm R}$$=$$1$,
$m_{\rm R}$$=$$0$
state 
are 
low-field 
seeking 
for fields of up 
to 
20~kV/cm,
and it is molecules in this state that
are used for our deflector.
} 
\end{figure}

The 
Stark
shift of the lowest electronic
and 
vibrational 
state of 
the 
BaF 
molecule has been 
discussed in 
Ref.~\cite{touwen2024manipulating},
and is calculated using methods 
first presented in 
Ref.~\cite{peter1957high}.
Fig.~\ref{fig:StarkShift}
shows the calculated 
energy levels of the lowest two
rotational states in the presence
of an electric field.
For fields of 
20~kV/cm or less, 
the 
$n_{\rm R}$$=$$1$,
$m_{\rm R}$$=$$0$
state
(here 
$n_{\rm R}$
is the rotational quantum number and 
$m_{\rm R}$
is its projection)
is a 
low-field-seeking 
state.

\begin{figure}[t!]
\centering
\includegraphics[width=3.0in]{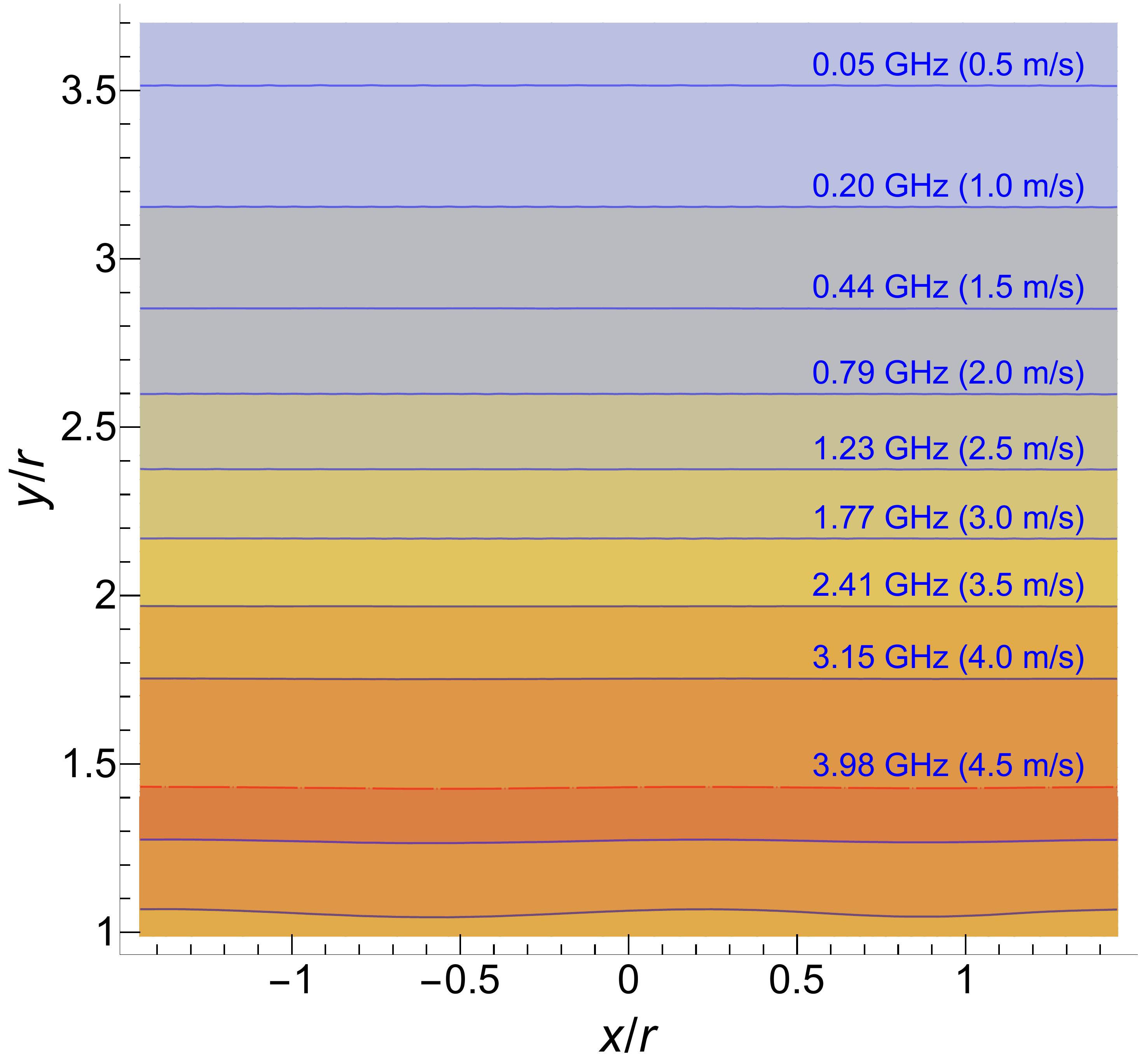}
\caption{
\label{fig:StarkPotential} 
(color online)
The 
potential experienced by a 
BaF 
molecule
in the 
$n_{\rm R}$$=$$1$,
$m_{\rm R}$$=$$0$
state,
which results from the 
Stark 
shift of 
Fig.~\ref{fig:StarkShift}
along with the electric 
field of 
Fig.~\ref{fig:VE}(b).
The potential on the electrodes
is tuned to have a field of 
20~kV/cm 
at a height of
$y$$\approx$$1.4r$.
The downward velocity of 
a 
BaF
molecule that can be reflected
by each contour is indicated.
} 
\end{figure}

Applying this 
Stark
shift to the field of 
Fig.~\ref{fig:VE}(b),
with the field set to 
20~kV/cm 
at a height of 
$y$$\approx$$1.4 r$
(where the contours of 
Fig.~\ref{fig:VE}(b)
become approximately flat),
leads to the potential
shown in 
Fig.~\ref{fig:StarkPotential}.
The strength of this potential is
sufficient to reflect
BaF 
molecules that have
a downward velocity component 
of
4.5~m/s
or less.

Note that the 
equipotential
surfaces are very nearly 
planar, 
being even more planar than 
the surfaces of 
Fig.~\ref{fig:VE}(b)
since the
20~kV/cm 
field is tuned to the maximum
shift,
as shown in 
Fig.~\ref{fig:StarkShift}.
Calculated trajectories show that 
reflections are nearly 
specular.
In particular, 
for molecules 
produced in a 
helium-buffer-gas 
source
(which would have a 
forward velocity of 
$>$100~m/s),
all calculated trajectories
have a difference between the
angle of incidence and
the angle of reflection
of less than one
microradian.
Even for much slower molecules
($\approx$5~m/s forward velocity),
which would be hard to produce
and
would lead to reflections that
are less grazing,
this difference would still be 
$\lesssim$10~microradians.

\section{\label{sec:deflector}Our design for separating B\lowercase{a}F molecules from other ablation products}

\begin{figure*}
\centering
\includegraphics[width=7in]{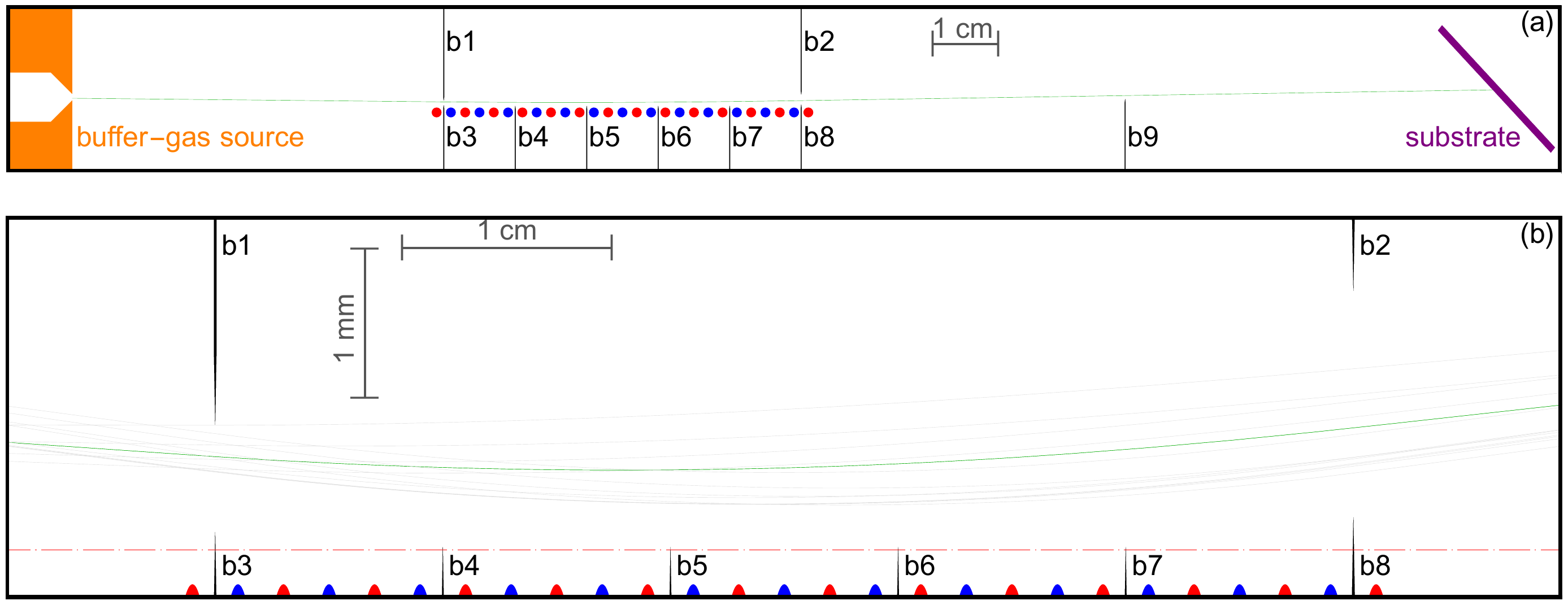}
\caption{
\label{fig:deflector} 
(color online)
The design for an electrostatic 
deflector for separating 
BaF molecules from
other 
laser-ablation 
products 
(a).
Panel
(b)
shows an 
expanded view
of the central
part of the apparatus,
with 
the horizontal 
and vertical
magnifications 
by factors of 
3
and
30,
respectively.
Several trajectories 
(green and gray)
are shown for 
BaF
molecules in the
$n_R=1$,
$m_R=0$
state.
The 
red 
dot-dash
line represents the plane from which
molecules with a downward velocity component of
4.5~m/s 
are reflected,
as also shown in 
Fig.~\ref{fig:StarkPotential}.
Blockers 
b1
through
b9
(which are razor sharp)
ensure that there are no 
straight-line
paths from the source to 
any part of the substrate.
Additionally, they block both spectral 
and 
nonspectral
reflection of molecules off 
the cylindrical electrodes
(red and blue) 
from reaching the substrate.
} 
\end{figure*}

Figure~\ref{fig:deflector}
shows our design that will be used to separate
BaF 
molecules
from other 
laser-ablation
products.
As our source uses 
ablated barium metal
in the presence of 
SF$_6$ 
to produce 
BaF
molecules 
in the 
buffer-gas cell,
many other atoms
(e.g., 
Ba),
molecules
(e.g., 
BaF$_2$,
SF$_6$)
and clusters
(e.g., 
Ba$_n$)
could be present.
Ions are easily separated
with a small electrostatic 
field. 
Our apparatus
\cite{corriveau2024matrix}
leaves a 
20-cm
spacing between the 
laser-ablation
buffer-gas 
source and the 
substrate on which 
the solid is grown.
This 
20-cm 
spacing is sufficient 
to allow small deflections 
possible from an electrostatic
deflector,
as described in this work,
(or similar small deflections
possible for 
laser-induced 
forces
\cite{
marsman2023large,
marsman2023deflection})
to separate 
the
BaF
molecules 
from the other
laser-ablation 
products.
The large spacing between 
the 
buffer-gas
source and the 
deflector 
and the efficient 
cryopumping 
by charcoal on the cryogenic 
surfaces
allows for a low
enough pressure 
near the deflector
to make the effects of 
collisions minimal.

The design
shown in 
Fig.~\ref{fig:deflector}
consists of a series of 
27
1.5-mm-diameter
stainless-steel 
cylinders that are 
alternately held at 
potentials of 
$-$2000~V
(red)
and 
$+$2000~V
(blue)
that create 
the required electric fields.
Two
razor-sharp
stainless-steel 
blockers
(b1
and
b9
in 
Fig.~\ref{fig:deflector})
ensure that no
atom, 
molecule,
or 
cluster
can travel directly from the 
1-mm-high
opening of the 
buffer-gas
source to any point 
on the 
sapphire substrate.
The sharp edges
of the blockers ensure
that 
BaF
molecules do not
make grazing reflections
off of the blockers themselves.
Additional blockers 
(b2 
through 
b8)
ensure that no reflections
(spectral 
or 
nonspectral)
off of the cylinders
can make it to the 
substrate.
The blockers 
are located in planes
where the potential
very nearly zero,
resulting in very 
small disturbances 
of the potential due 
to their presence.

A typical trajectory for 
a deflected 
BaF
molecule
(in the 
$n_{\rm R}$$=$$1$,
$m_{\rm R}$$=$$0$
state)
is shown in 
panel~(a)
of the figure.
Panel~(b)
shows an expanded view
(expanded by a factor of
3
in the horizontal direction 
and a factor of 
30
in the vertical direction)
with a larger set of trajectories.
Panel 
(b) 
also shows the surface 
of reflection for 
BaF
molecules with a downward 
component of
velocity of 
4.5~m/s,
represented by  
the dot-dashed line 
(as in  
Fig.~\ref{fig:StarkPotential}).

A simulation of this deflection 
is performed 
for a thermal source of molecules
with a 
4-kelvin 
temperature
and an average forward 
speed of 
150~m/s
(the typical speed for 
a helium 
buffer-gas
source
\cite{truppe2018buffer}).
In the simulation,
molecules start 
from the 
1-mm-high
output aperture 
of the 
buffer-gas 
source.
Most molecules are 
intercepted by one of the 
blockers 
(b1
through
b9
in 
Fig.~\ref{fig:deflector}),
but molecules in the 
$n_{\rm R}$$=$$1$,
$m_{\rm R}$$=$$0$
state
that have initial 
downward velocity components of 
0.1 
to
3~m/s
can be deflected onto
the central 
one-millimeter-high
band
of the substrate.
The simulation shows that 
the 
flux of 
$n_{\rm R}$$=$$1$,
$m_{\rm R}$$=$$0$
molecules that 
are deflected to this 
band on the substrate 
is equal to 
62\% 
of the 
flux that 
would have been incident
on the 
same area if there were 
no deflector present.
Thus the deflector design
is expected to  
efficiently deflect 
BaF molecules.
No other atom, 
molecule or clusters
created in the 
ablation source
is expected to 
have a large enough permanent
dipole moment or 
polarizability
to be significantly deflected 
by the electric fields 
in this deflector.

To realize the 
62\% deflection efficiency,
the 
BaF 
molecules would have to be
optically pumped
into the 
$n_{\rm R}$$=$$1$,
$m_{\rm R}$$=$$0$
state.
At 
4~K,
more than 
70\%
of the 
molecules will be in the 
$n_{\rm R}$$=$$0$,
$1$,
$2$
and
$3$
states.
A laser driving transitions from the 
$X\,^2\Sigma_{1/2}(v$$=$$0$,$\,n_{\rm R}$$=$$3)$
state
to the 
$A\,^2\Pi_{1/2}(v$$=$$0$,$\,j$$=$$3/2)$
positive-parity
state
will move the 
$n_{\rm R}$$=$$3$
population into the 
$n_{\rm R}$$=$$1$
state.
Two other lasers tuned
to drive transitions from 
the
$X\,^2\Sigma_{1/2}(v$$=$$0)$
states with
$n_{\rm R}$$=$$0$
and 
$2$
to the 
$A\,^2\Pi_{3/2}(v$$=$$0$,$\,j$$=$$3/2)$
state
will drive 
the 
$n_{\rm R}$$=$$0$
and 
$2$
populations 
into the 
$X\,^2\Sigma_{1/2}(v$$=$$0$,$\,n_{\rm R}$$=$$1)$
state.
The 
$A\,^2\Pi_{3/2}$
state is used since 
its very small 
lambda-doubling 
separation allows 
for mixing of 
odd- 
and 
even-parity
states in 
even very small 
electric fields,
allowing for decays to 
both
odd-
and 
even-parity 
states.
A final laser beam 
(downstream of the others),
this one in an applied electric
field,
is tuned to the transition from the 
$X\,^2\Sigma_{1/2}(v$$=$$0$,$n_{\rm R}$$=$$1$,$|m_{\rm R}|$$=$$1)$
state
to the 
$A\,^2\Pi_{1/2}(v$$=$$0$,$j$$=$$1/2)$
positive-parity
state to
move the population into the 
$X\,^2\Sigma_{1/2}(v$$=$$0$,$n_{\rm R}$$=$$1$,$m_{\rm R}$$=$$0)$
state. 
The losses to 
higher-$v$
states is small 
due to the
nearly
vertical 
transitions
from the 
$X\,^2\Sigma_{1/2}(v$$=$$0)$
state to 
the 
$A\,^2\Pi_{1/2}(v$$=$$0)$
state,
with a branching 
ratio of 
3.5\%
\cite{hao2019high})
for decay to 
higher-$v$
levels.

\section{\label{sec:concl}
Conclusions}

An electrostatic mirror with
nearly spectral reflections is 
described. 
A design for a deflector of 
BaF
molecules,
which will separate these 
molecules from other 
laser-ablation products is 
presented.

\section{\label{sec:ackn}
Acknowledgements}

We acknowledge support from
the 
Gordon and Betty Moore Foundation,
the
Alfred P. Sloan Foundation,
the
John Templeton Foundation 
(through 
the 
Center for Fundamental Physics
at
Northwestern University),
the 
Natural Sciences
and Engineering Council 
of Canada, 
the 
Canada Foundation for Innovation, 
the
Ontario Research Fund
and
from
York University.

\bibliography{polarbeam}

\end{document}